\documentstyle[12pt]{article}
\newcommand{\alt}{\mathrel{\raisebox{-.6ex}{$\stackrel{\textstyle<}{\sim}$}}}

\def\overlay#1#2{\ifmmode \setbox 0=\hbox {$#1$}\setbox 1=\hbox to\wd 0{\hss
$#2$\hss }\else \setbox 0=\hbox {#1}\setbox 1=\hbox to\wd 0{\hss #2\hss }\fi
#1\hskip -\wd 0\box 1}
\catcode`@=11
\newcount\@tempcntc
\def\@citex[#1]#2{\if@filesw\immediate\write\@auxout{\string\citation{#2}}\fi
  \@tempcnta\z@\@tempcntb\m@ne\def\@citea{}\@cite{\@for\@citeb:=#2\do
    {\@ifundefined
       {b@\@citeb}{\@citeo\@tempcntb\m@ne\@citea\def\@citea{,}{\bf ?}\@warning
       {Citation `\@citeb' on page \thepage \space undefined}}%
    {\setbox\z@\hbox{\global\@tempcntc0\csname b@\@citeb\endcsname\relax}%
     \ifnum\@tempcntc=\z@ \@citeo\@tempcntb\m@ne
       \@citea\def\@citea{,}\hbox{\csname b@\@citeb\endcsname}%
     \else
      \advance\@tempcntb\@ne
      \ifnum\@tempcntb=\@tempcntc
      \else\advance\@tempcntb\m@ne\@citeo
      \@tempcnta\@tempcntc\@tempcntb\@tempcntc\fi\fi}}\@citeo}{#1}}
\def\@citeo{\ifnum\@tempcnta>\@tempcntb\else\@citea\def\@citea{,}%
  \ifnum\@tempcnta=\@tempcntb\the\@tempcnta\else
   {\advance\@tempcnta\@ne\ifnum\@tempcnta=\@tempcntb \else \def\@citea{--}\fi
    \advance\@tempcnta\m@ne\the\@tempcnta\@citea\the\@tempcntb}\fi\fi}
\catcode`@=12
\catcode`@=11
\newtoks\@stequation
\def\mathletters{\refstepcounter{equation}%
  \edef\@savedequation{\the\c@equation}%
  \@stequation=\expandafter{\theequation}
  \edef\@savedtheequation{\the\@stequation}
  \edef\oldtheequation{\theequation}%
  \setcounter{equation}{0}%
  \def\theequation{\oldtheequation\alph{equation}}}
\def\endmathletters{%
  \setcounter{equation}{\@savedequation}%
  \@stequation=\expandafter{\@savedtheequation}%
  \edef\theequation{\the\@stequation}%
  \global\@ignoretrue}
\catcode`=12

\begin{document}

\font\fortssbx=cmssbx10 scaled \magstep2
\hbox to \hsize{
\hskip.5in \raise.1in\hbox{\fortssbx University of Wisconsin - Madison}
\hfill\vtop{\hbox{\bf MAD/PH/779}
                \hbox{\bf UCD-94-12}
                \hbox{\bf NUHEP-TH-94-8}
                \hbox{April 1994}}}

\vspace{.5in}

\begin{center}
{\bf Production of Weak Bosons and Higgs Bosons in \boldmath{$e^-e^-$}
Collisions}\\[.4in]
V.~Barger$^{a}$, J.F.~Beacom$^a$, Kingman Cheung$^b$, and T.~Han$^c$\\[.2in]
\it $^a$Physics Department, University of Wisconsin, Madison, WI 53706\\
$^b$Department of Physics \& Astronomy, Northwestern University,
Evanston, IL 60208\\
$^c$Department of Physics, University of California, Davis, CA 95616
\end{center}

\vspace{.5in}

\begin{abstract}
We present calculations of cross sections for single $W^-$ and $Z$
production and $W^-W^-,\ W^-Z,\ W^-\gamma,\ ZZ$, and $W^+W^-$ pair
production within the Standard Model at $e^-e^-$ linear colliders.  We
evaluate Standard-Model Higgs boson production in the channels $e^-e^-
\to e^-e^- H$, $e^-\nu W^- H$, and $e^-e^- ZH$.  We also illustrate
the enhancements in the $W^-W^-$ cross section that would result from
a strongly-interacting Higgs sector or from a $H^{--}$ resonance in a
Higgs doublet + triplet model.
\end{abstract}

\thispagestyle{empty}

\newpage

\section{Introduction}

High-energy experiments at $e^+e^-$ colliders have proved to be very
fruitful.  The measurements at LEP\,I and SLC give precision tests of
the Standard Model for electroweak interactions\cite{LEP}. In
addition, interesting direct limits have been placed on the Higgs
boson mass and the masses of supersymmetric and other new
particles\cite{newlimits}. As a means to continue to address these
important issues, the design of linear $e^+e^-$ colliders with
center-of-mass energies in the range 0.3--2.0~TeV is being actively
pursued worldwide\cite{hawaii}. With an appropriate design, an
$e^+e^-$ collider could also be operated as an $e^-e^-$
collider\cite{oldslc}.

The construction of high energy $e^-e^-$ colliders has not been
pursued in recent times. The probable reason for this lack of activity
is the absence of $s$-channel resonance production in $e^-e^-$ due to
lepton number conservation. However, precisely because direct channel
resonances are not expected, high energy $e^-e^-$ collisions could be
a clean way to uncover physics beyond the Standard Model (SM), and
several new physics signals have recently been discussed\cite{heusch}.

There are a number of interesting physics possibilities for an
$e^-e^-$ collider.  In the context of supersymmetry, searches could be
made for scalar electron and chargino pair
production\cite{cuypers}. For example, the chargino production process
$e^-e^-\to \chi^-\chi^-$ with $\chi^-\to W^-\chi^0$ decays gives
$W^-W^-$ final states. Searches could also be made for doubly-charged
Higgs bosons\cite{georgi,dicus,gunion2,chiv}, doubly-charged gauge
bosons\cite{frampton} or $W_R$ production\cite{rizzo}. Massive
Majorana neutrino exchange could allow $e^-e^-\to W^-W^-$
production\cite{rizzo3}, although due to double $\beta$-decay
constraints the cross section must be very small.  Tests of anomalous
vector boson couplings could be made\cite{choud}, such as the
$WW\gamma$ coupling from the $e^-e^-\to e^-W^-\nu_e$ process.

In the present paper we calculate SM expectations for single and pair
production of vector bosons in $e^-e^-$ collisions which are of
interest for tests of electroweak theory and as backgrounds to new
physics possibilities. We calculate the associated production of a
neutral Higgs boson and a weak boson.  We also study the possibility
of detecting strong $W^-W^-$ scattering and quantitatively evaluate
the enhancement of $W^-W^-$ production due to a doubly-charged Higgs
boson.
At the end, we will briefly discuss the effects of beamstrahlung and
bremsstrahlung on the effective center-of-mass energies and the
effective luminosities of the $e^-e^-$ colliders. We also consider the
advantages of having both $e^-$ beams polarized.

\section{Single weak boson production}

Production of weak bosons occurs at order $\alpha^3$ in the processes
\begin{mathletters}
\begin{eqnarray}
e^-e^- &\to& e^-W^-\nu_e \;, \label{eq:a^3 a}\\ e^-e^- &\to& e^- Z e^-
\;, \label{eq:a^3 b}
\end{eqnarray}
\end{mathletters}
whose Feynman diagrams are given in Fig.~\ref{fig:a^3feyn}. The
helicity amplitudes can be straightforwardly written down.  The
resulting total cross sections are shown versus $\sqrt s$ in
Fig.~\ref{fig:a^3tot}.  At $\sqrt s=0.5$~TeV the $W^-$ and $Z$ cross
sections are approximately equal and have the value 7~pb; thus with
10~fb$^{-1}$ luminosity about $7\times 10^4$ each of single $W$ and
single $Z$ events would be produced at this energy.

In typical accelerator designs the beam hole region is
$\theta<15^\circ$ (or equivalently the beam hole pseudorapidity range
is $|\eta|>2$). We have evaluated the effect of making such an
acceptance cut on the leptons ($e^-, \mu^-$) or quark jets from $W$
and $Z$ decays, including the full spin correlations from the weak
boson production and decay sequence. With the lepton (quark) decay
products of the weak bosons within the $15^\circ< \theta < 165^\circ$
acceptance region the cross sections at $\sqrt s=0.5$~TeV are 1.4
(4.0)~pb for $W$ and 0.15 (1.6 )~pb for $Z$, including the decay
branching fractions.

\section{Double weak boson production}

Pair production of vector bosons occurs at order $\alpha^4$ in the
processes
\begin{mathletters}
\begin{eqnarray}
e^-e^- &\to& e^-e^- W^+W^- \;, \label{eq:a^4 a}\\ e^-e^- &\to&
e^-\nu_e W^-Z \;, \label{eq:a^4 b}\\ e^-e^- &\to& \nu\nu W^-W^- \;,
\label{eq:a^4 c}\\ e^-e^- &\to& e^-\nu_e W^- \gamma \label{eq:a^4 d}
\\ e^-e^-&\to& e^-e^- ZZ \;. \label{eq:a^4 e}
\end{eqnarray}
\end{mathletters}
The Feynman diagrams for reaction (\ref{eq:a^4 c}) are shown in
Fig.~\ref{fig:a^4feyn}.

The analogues of most of these processes at $e^+e^-$ colliders have
been calculated in Ref.~\cite{bckp}.  The formulas for the $e^+e^-$
amplitudes given in Ref.~\cite{bckp} can be properly crossed
to obtain the corresponding amplitudes for $e^-e^-$ collisions.
The
$e^+e^-$ analogue of the process (\ref{eq:a^4 d}) was not previously
considered but these amplitudes can be obtained by modifying the
$e^-e^-\to e^-\nu_e W^-Z$ amplitudes, changing the $Z$-couplings to
the corresponding $\gamma$ couplings.

Since the process $e^-e^-\to e^-\nu_e W^-\gamma$ is essentially the
photon-bremsstrahlung contribution to the single $W$ production, we
impose acceptance cuts on the final state $\gamma$ to distinguish it
from the single $W$ process.  The chosen acceptance cuts are
\begin{equation}
p_{T\gamma}>10\;{\rm GeV},\quad |\cos\theta_\gamma|<\cos(15^\circ),
\quad
\theta(\gamma,e^-_{\rm final})>5^\circ\,,
\end{equation}
where $\theta(\gamma,e^-_{\rm final})$ represents the angle between
the photon and the final state electron 3-momenta.

Figure~\ref{fig:a^4tot} shows the total cross sections for the
reactions (\ref{eq:a^4 a})--(\ref{eq:a^4 e}) versus $\sqrt s$ for
$m_H=100$~GeV.  The value of $m_H=100$~GeV is taken as representative
since the cross section is insensitive to the value of $m_H$ for
$m_H\alt 2M_W$.  The large size of the $W^+W^-,\ W^-\gamma,$ and
$W^-Z$ processes is due to photon exchange contributions.

\section{Standard-Model Higgs production}

In $e^+e^-$ collisions the Standard-Model Higgs boson is produced via
$e^+e^-\to\bar\nu_e\nu_e H$ and $e^+e^-\to ZH$ processes. In $e^-e^-$
collisions single Higgs production can occur via the processes
\begin{mathletters}
\begin{eqnarray}
e^-e^- &\to& e^-e^-H \label{eq:sm a}\\ e^-e^- &\to& e^-\nu_e W^- H
\label{eq:sm b}\\ e^-e^- &\to& e^-e^-ZH \label{eq:sm c}
\end{eqnarray}
\end{mathletters}
The tree-level Feynman diagrams for (\ref{eq:sm a}) and (\ref{eq:sm
b}) are shown in Fig.~\ref{fig:smfeyn}.  The cross sections are shown
in Fig.~\ref{fig:smtot} versus $\sqrt s$ for $m_H=100$~GeV and versus
$m_H$ at $\sqrt s=0.5$~TeV; for $m_H=100$~GeV the $e^-e^-\to e^-e^-H$
cross section is 9~fb. In comparison to the Higgs production cross
section at $\sqrt s=0.5$~TeV in $e^+e^-$ collisions are 60~fb for
$Z^*\to ZH$ and 95~fb for $W^*W^*\to H$ mechanisms \cite{bckp}.  Higgs
production can also occur via $\gamma\gamma$ and $\gamma Z$ fusion at
the one-loop level.

For $m_H$ close to $M_Z$ the $Z$-production processes (\ref{eq:a^3
b}), (\ref{eq:a^4 b}), (\ref{eq:a^4 e}) are large backgrounds to the
Higgs processes (\ref{eq:sm a}), (\ref{eq:sm b}), (\ref{eq:sm c})
respectively.  For example, for the production of a 100~GeV Higgs
boson at $\sqrt s=0.5$~TeV, $\sigma(e^-e^-\to e^-e^-H)=9$~fb and
$\sigma(e^-e^-\to e^-e^-Z)=7000$~fb. The Higgs boson decays dominantly
to $b\bar b$ while the $Z\to b\bar b$ branching fraction is 0.15. The
$H$-production is central while the $Z$-production is peaked at
forward and backward scattering angles; thus the $Z$ background can be
selectively suppressed by angular cuts.  In the following we consider
the observability of $e^-e^-\to e^-e^- H$ with $H\to b\bar b$ decays.

\begin{table}
\centering
\caption[]{
\label{table1}
\small
The Higgs boson signal in the production of $e^-e^-\to e^-e^- H$ for
$m_H=60$--$140$ GeV and the background from $e^-e^-\to e^-e^- b\bar b$
with the invariant mass of the $b\bar b$ pair in the range $m_H\pm
\Delta m_H$, with $\Delta m_H=10$~GeV.  For simplicity we take B$(H\to
b\bar b)=1$, and assume that all the signal falls within $m_H\pm\Delta m_H$.
The acceptance cuts are $p_{Te} > 15$ GeV and $|\cos\theta_e| <
\cos (15^\circ)$ on the final state electrons, and $p_{Tb} > 25$ GeV
and $|\cos\theta_b| < 0.7$ on the final state $b$'s.  }
\medskip
\begin{tabular}{|c|c|c|}
\hline
$m_H$     &   Signal (fb)   &  Background (fb) \\
\hline
\hline
60        &     1.4         & 0.02  \\
70        &     1.4         & 0.03  \\
80        &     1.4         & 0.27   \\
90        &     1.4         & 1.1    \\
100       &     1.4         & 0.86    \\
120       &     1.3         & 0.02    \\
140       &     1.3         & 0.01   \\
\hline
\end{tabular}
\end{table}

The complete $e^-e^-\to e^- e^- b \bar b$ background includes both
$e^-e^-\to e^-e^-Z$ with $Z\to b\bar b$ and the two-photon production
of $b\bar b$.  Based on the fact that the two-photon background can be
reduced substantially by keeping the photon propagators far off-shell,
we impose the following acceptance cuts
\begin{equation}
p_{Te} > 15\;{\rm GeV} \qquad {\rm and} \qquad
|\cos\theta_e| < \cos (15^\circ) \;, \label{pTe cut}
\end{equation}
on both of the electrons in the final state.  We also impose the
following acceptance cuts on the $b$'s in the final state:
\begin{equation}
p_{Tb} > 25\;{\rm GeV} \qquad {\rm and} \qquad  |\cos\theta_b| < 0.7 \;.
\label{pTb cut}
\end{equation}
The latter cuts are tailored to reduce the $eeZ$ background, as the
$b$'s from $Z$-boson production are much more forward-peaked than the
$b$'s from Higgs-boson production.

After imposing (\ref{pTe cut}) and (\ref{pTb cut}), the Higgs signal is
1.4~fb for $m_H=100$~GeV while the total $eeb\bar b$ background
integrated over all $m(b\bar b)$ is 1.3~fb.  The background has a
wide $m(b\bar b)$ invariant mass distribution (see
Fig.~\ref{fig7}) with a peak at $m(b\bar b)=M_Z$ due to $e^-e^-\to
e^-e^-Z$.  The signal is a sharp peak at the Higgs mass.  We consider
an invariant mass resolution $\Delta m_H$ for the $b\bar b$ pair of 10
GeV and assume that all the signal falls within $m_H\pm \Delta m_H$.
The signal and the background in such bins at various Higgs mass
values are summarized in Table~\ref{table1}.  In this simple
comparison we have assumed perfect $b$-tagging with only the $b\bar b$
final state counted as background, {\it i.e.}, the other $q\bar
q\;(q=u,d,s,c)$ final states are rejected.  The background for
$m_H=90$ GeV is the largest because of $ee\to eeZ$.  We conclude
that the intermediate mass Higgs boson in the channel $e^-e^-\to e^-
e^- H$ with $H\to b\bar b$ may be observable with an integrated
luminosity of at least 20 fb$^{-1}$.

When $m_H>2m_Z$,  Higgs
production with $H\to ZZ$ decay can give an appreciable enhancement to
the $e^-e^-\to e^-e^-ZZ$ production; for example, at $m_H=200$~GeV the
cross section of $e^-e^-\to e^-e^-ZZ$ is a factor of two larger than
that with $m_H=100$~GeV.

\section{Strong $W_LW_L$  Scattering}

If no light Higgs boson is found for $m_H^{} \alt 800$ GeV, one would
anticipate that the interactions among longitudinal vector bosons
become strong\cite{chan-gail}. An $e^-e^-$ collider offers a unique
opportunity to explore the weak isospin $I=2$ $s$-channel \cite{wpwp}
via the process $W_L^-W_L^- \to W_L^-W_L^-$ \cite{hantalk}.  The
simplest model for a strongly-interacting $W_L^-W_L^-$ sector is the
exchange of a heavy Higgs boson. This results in an enhancement of the
$e^-e^-\to\nu\nu W^-W^-$ production cross section compared to that
expected from the exchange of a light Higgs boson. This enhancement
due to a Higgs boson of mass 1~TeV can be defined as the difference of
the $W_L^-W_L^-\to W_L^-W_L^-$ fusion contributions
\begin{equation}
\Delta\sigma_H = \sigma(m_H=1~{\rm TeV}) - \sigma(m_H=0.1~\rm TeV)
\end{equation}
to $e^-e^-\to\nu\nu W^-W^-$ production. There is no appreciable
numerical change between the choices $m_H=0.1$~TeV and $m_H=0$ for the
light Higgs boson reference mass. We find the values
\begin{equation}
 \Delta \sigma_H \simeq \left\{
\begin{array}{ll}
53.6 - 50.9 = 2.7~{\rm fb}
& \mbox{at $\sqrt s = 1.5$~{\rm TeV}; } \\
86.5 - 82.0 = 4.5 ~{\rm fb}
& \mbox{at $\sqrt s = 2$~{\rm TeV}. }
\end{array}
\right.
\end{equation}

In the following we address the observability of this strong
$W_L^-W_L^-\to W_L^-W_L^-$ signal at $\sqrt s=2$~TeV. The complete
Standard-Model cross section for $e^-e^-\to\nu\nu W^-W^-$ from all
diagrams in Fig.~\ref{fig:a^4feyn} is about 20 times larger than
$\Delta\sigma$. Hence the background contributions associated with
transverse $W$ bosons ($W_T^-W_T^-$, $W_T^-W_L^-$) must somehow be
selectively reduced by acceptance criteria if we are to observe the
strongly-interacting $W_L^-W_L^-$ signal.

There are several different ways to accomplish the substantial
background suppression\cite{hantalk,baggeretal,cheung}.  The
$W_L^-W_L^-$ scattering process gives large $M(W^-W^-)$ of order 1~TeV
with centrally-produced $W^-$ having large $p_T$. Thus we impose the
kinematic cuts
\begin{eqnarray}
 p_T(W) > 150 ~{\rm GeV},  \quad  |\cos \theta_W| < 0.8 \label{pT(W)} \,,
\end{eqnarray}
which retains about one-third of the signal and reduces the SM
backgrounds by more than an order of magnitude.  After those cuts, the
heavy Higgs enhancement becomes $\Delta \sigma_H \simeq 7.8 - 6.3 =
1.5$~fb.

In hadronic $W$-decays, the sign of the $W$ charge is not identified
and the two-photon process $e^-e^-\to e^-e^-W^+W^-$ may also present a
substantial background when the final electrons are not observed. From
Fig.~\ref{fig:a^4tot} the cross section for $e^-e^-\to e^-e^-W^+W^-$
is about a factor of 30 larger than $e^-e^-\to \nu\nu
W^-W^-$. Moreover, $e^-e^-\to e^-\nu W^-Z$ could add to the background
if the reconstructed invariant masses in hadronic decays are not
sufficiently resolved to distinguish $W$ from $Z$. In order to
suppress these backgrounds we veto events in which an electron can be
identified having\cite{hagiw}
\begin{eqnarray}
 E_e > 50 ~{\rm GeV},  \quad  |\cos \theta_e| < |\cos(150~{\rm mrad})| \,.
\label{E_e}
\end{eqnarray}
With the acceptance of Eqs.~(\ref{pT(W)}) and (\ref{E_e}), the
remaining $e^-e^-\to e^-e^-W^+W^-$ and $e^-e^-\to e^-\nu_e W^-Z$
backgrounds are 60 fb and 10 fb, respectively.

A further improvement in isolating the $W_L^-W_L^-$ signal derives
from the fact that the $p_T(WW)$ spectrum of the signal is peaked
around $M_W$ and falls off rapidly at high $p_T$ like $1/p_T^4$.
Figure~\ref{fig:pT}(a) compares the $p_T(VV)$ distribution of the
$W_L^-W_L^-$ signal with the backgrounds. Note that the difference
between the solid curve (with $m_H=1$~TeV) and the dashed curve
(with $m_H=0.1$~TeV) is the strong $W_L^-W_L^-$ enhancement. We impose
the selection
\begin{equation}
50~{\rm GeV} < p_T(VV) < 300~\rm GeV \label{pT(VV)}
\end{equation}
for additional background suppression.

The signal gives $W_L^-$ bosons that are fast and moving back-to-back
in the transverse plane. The difference in the transverse momenta of
the two weak bosons is
\begin{equation}
\Delta p_T(VV) = |{\bf p}_T(V_1)-{\bf p}_T(V_2)|
\end{equation}
presented in Fig.~\ref{fig:pT}(b). The signal (difference of solid and
dashed curves) is enhanced by the cut
\begin{equation}
\Delta p_T (VV) > 400\ \rm GeV \,. \label{Delta pT}
\end{equation}
With the additional cuts of Eqs.~(\ref{pT(VV)}) and (\ref{Delta pT})
the surviving signal is
\begin{equation}
\Delta \sigma_H \simeq 3.8 - 2.8 = 1.0 \ \rm fb \,.
\end{equation}
The efficiency for retaining the signal is 67\%. The remaining
backgrounds are 4.4~fb for $e^-e^-\to e^-e^- W^-W^+$ and 4.7~fb for
$e^-e^-\to e^-\nu W^-Z$.
The resulting $M(VV)$ distributions after these cuts are presented in
Fig.~\ref{fig:invar}. At high $VV$ invariant masses the strong
$W_L^-W_L^-$ scattering rate due to the exchange of a 1~TeV Higgs
boson is enhanced over the $W^-_TW^-_T$, $W_T^-W_L^-$ and $W^-W^+$ backgrounds,
while the background due to $W^-Z$ still persists.

Next we estimate the signal rates for other strongly-interacting
scenarios.  We consider a chirally-coupled scalar boson ($m_S=1$ TeV
and $\Gamma_S=350$ GeV), a chirally-coupled vector boson ($m_V=1$ TeV
and $\Gamma_V=25$ GeV)\cite{hantalk,baggeretal}, and the low energy
theorem amplitude\cite{baggeretal,LET}.  These calculations are
carried out with the effective $W$-boson approximation (EWA)
\cite{EWA}. In this calculational method one is unable to obtain the
exact kinematics for the final state of $W^-W^-$.  In order to
simulate the acceptance effects of Eqs.~(\ref{pT(VV)}) and (\ref{Delta
pT}), we multiplied the EWA calculations by the efficiency factor 67\%
found in the heavy Higgs boson model.

The predicted cross sections at $\sqrt s = 2$~TeV with the cuts discussed
above are presented in Table~\ref{table2}. Also given in parentheses
are the number of events with hadronic $W,Z$ decays for an integrated
luminosity of 300~fb$^{-1}$.  In Table~\ref{table2} we see that the
backgrounds (dominantly from $W^-Z$ production) to the signals are
still substantial after the kinematic selection criteria.  Due to the
absence of an $s$-channel resonance, the signals are mostly an overall
enhancement on the $M_{WW}$ spectrum.  If we can predict the SM
backgrounds at a level of better than 10\%, there is a chance that we
can observe the strong $W^-W^-$ scattering
via the hadronic decay modes at statistical significance
$S/\sqrt B > 4$ for a 1~TeV scalar or a vector particle, and at
$S/\sqrt B \geq 6.4$ for the LET amplitude with $M_{WW} > 500$~GeV.

\begin{table}
\centering
\caption[]{
\label{table2}
\small  Cross sections  at $\sqrt s=2$~TeV from different
models of strongly-interacting $W^-W^-$ with cuts discussed in the
text.  Backgrounds are summed over $W^-W^-$ with a light Higgs
exchange, $W^+W^-$, and $W^-Z$. Entries are in units of fb and those
in parentheses correspond to the number of events with hadronic $W,Z$
decays for an integrated luminosity of 300 fb$^{-1}$.}
\medskip
\begin{tabular}{|l|c|c|c|c|c|c|} \hline
$M_{WW}^{min}$ & SM  & Scalar & Vector   & LET & Backgrounds \\
\noalign{\vskip-1ex}
& $m_H=1$ TeV & $m_S=1$ TeV & $m_V=1$ TeV & & \\
\hline
0.5 TeV
& 0.88 (130)  & 1.2 (175)  & 1.1 (167) & 1.7 (245)  &  10 (1470) \\ \hline
0.75 TeV
&  0.44 (65) & 0.72 (106) & 0.63 (93) & 1.0 (150)  & 3.5 (515)\\ \hline
1 TeV
& 0.15 (22)   & 0.31 (46)  & 0.26 (38) & 0.48 (71) & 1.0 (147) \\ \hline
\end{tabular}
\end{table}

If separation of $W$ and $Z$ peaks can be achieved in the $m_{jj}$
mass distributions then the backgrounds would be significantly further
reduced.  The use of $Z \to e^+ e^-, \mu^+\mu^-, b \bar b$ (with
$b$-tagging), with combined branching fraction of about 22\%, could be
helpful in determining the contribution of the $W^-Z$ process.

\section{$H^{--}$ Signal in Scalar Doublet + Triplet Model}

The charged Higgs boson in SUSY and other two-Higgs-doublet models can
be pair-produced~\cite{rizzo2} via $e^-e^- \to \nu \nu H^-H^-$ via
the $W^-W^-$ fusion subprocess with $H^0$ exchange; see
Fig.~\ref{fig:H--}(a).  If the $H^-$ is not degenerate in mass with
$W$, this $H^-H^-$ signal could be spectacular and sensitive to the
model parameters, such as the ratio of vacuum expectation values
$\tan\beta=v_2/v_1$.

A search for a doubly-charged Higgs boson in a model with a scalar
triplet could also be carried out at an $e^-e^-$ collider.  What makes
the $e^-e^-$ collider unique in this instance is that the
doubly-charged Higgs boson can be produced as an $s$-channel resonance via
$W^-W^- \to H^{--}$ [see Fig.~\ref{fig:H--}(b)], followed by $H^{--}
\to W^-W^-, W^-H^-$ and possibly $f \bar f'$ decays \cite{gunion2},
depending on the model parameters. Detailed analyses of the doublet
plus triplet model can be found in the literature\cite{gunion2} and we
will not repeat the discussion of the model here.  The model contains
an SU(2) five-plet $H_5^{++,+,0,-,--}$, a triplet $H_3^{+,0,-}$, and 2
singlets $H_1^0$ and $H_1^{0'}$.  For simplicity, we assume that in
the Higgs potential\cite{gunion2} that there is no mixing
between the singlets (therefore $\lambda_3=0$) and that all other
$\lambda$-couplings are of similar size ($\lambda_1 \simeq \lambda_2
\simeq \lambda_4 \simeq \lambda_5$).
There are then two independent parameters left in the model: the mass
parameter of the five-plet, $m_5$, and the mixing angle $\theta_H$
between the Higgs doublet and the triplet fields, which is related to
the ratio of the vacuum expectation values.
Figure~\ref{fig:MWWdoubly} presents the $M_{W^-W^-}$ distribution at
$\sqrt s=0.5$~TeV including the resonance process $e^-e^- \to
\nu \nu H^{--} \to \nu \nu W^-W^-$ with $m_5=M(H^{--})=0.2$ or
0.3~GeV, taking maximum mixing $\tan\theta_H=1$.  A significant
enhancement above the SM background occurs.

\section{Beamstrahlung Effects}

Since beamstrahlung is a collective phenomenon, it depends very much
on the design of the machine, {\it e.g.}, the number of particles in a
bunch, the number of bunches in the beam, and the shape of the beam.
Systematic studies of beamstrahlung at $e^+e^-$ colliders have been
made; see Ref.~\cite{chen}.  Based on these analyses we briefly
comment on beamstrahlung effects on the luminosity and the
center-of-mass energy of an $e^-e^-$ collider.

In $e^+e^-$ collisions, the forces on the charged particles in one of
the beams due to the electromagnetic field of the opposite beam cause the
particles to curve towards the center of the beam, resulting in a more
compact beam and so the effective luminosity increases.  For example,
in the NLC design\cite{hawaii} the luminosity enhancement factor $H_D$ is
1.4, while that for the DLC design\cite{hawaii} is 2.8.  The effect of
beamstrahlung on the luminosities in $e^-e^-$ collisions is exactly
opposite, due to the fact that the force from the opposite beam causes
the particle to diverge from the center of the beam.  The decrease in
the luminosity of the $e^-e^-$ colliders (assuming the same design as
the corresponding $e^+e^-$ colliders) is a factor of 0.7 for the NLC
design and 0.36 for the DLC design.  However, the luminosity decreases
can likely be reduced by the altering the designs. Bremsstrahlung does
not change the luminosity.

The effect of beamstrahlung on the center-of-mass energy $\sqrt{s}$ is
easy to understand.  The forces felt by one of the beams due to the
fields of the opposite beam in either $e^+e^-$ or $e^-e^-$ collisions
are equal, though in opposite directions.  The accelerations of the
charged particles in the beam are therefore equal for $e^+e^-$ and
$e^-e^-$ collisions (again in opposite directions), and so the
radiations from the accelerated charges are the same.  Thus, the
resultant decreases in the center-of-mass energies due to radiation
are the same for $e^+e^-$ and $e^-e^-$ colliders.  In the DLC, NLC,
JLC, and TESLA designs\cite{hawaii} the decreases in the
center-of-mass energies due to beamstrahlung effects can be controlled
better than that of bremsstrahlung, which is well understood.  Since
the cross sections of all the processes considered in this paper
increase with $\sqrt{s}$, the decrease in the effective center-of-mass
energy due to beamstrahlung and bremsstrahlung will decrease the
cross sections.

\section{Advantages of Polarizing the Electron Beams}

There are advantages of polarizing the colliding electron beams either
left-handed or right-handed.  Only the left-handed fermions
($e^-_L,\,\nu_L,\,u_L,\,d_L,\dots$) and the right-handed anti-fermions
($e^+_R,\,\bar \nu_R,\, \bar u_R,\, \bar d_R,\dots$) couple to the SM
$W$ boson.  By the use of polarized electrons the backgrounds can be
reduced according to the nature of the signals of interest.  For
example, to search for the right-handed $W$ boson that exists in
left-right models, it will be advantageous to use $e^-_R$ beams in
order to reduce the SM  $W$ production.  By colliding $e_L^-$
beams the strong $W^-_L W^-_L$ signal in the process $e^-e^-\to \nu_e
\nu_e W^-_L W^-_L$ (where $W_L$ denotes longitudinal polarization)
can be enhanced over the $W^+W^-$, $ZZ$, and $W^-Z$
backgrounds.  Similarly, the signals of $H^{--}$ and $H^-H^-$
will be favored with the use of $e^-_L$ beams.  In
Table~3 we compare the relative sizes of the polarized cross sections
for the SM processes that we have calculated in this paper.  There are
four possible polarization combinations (LL, LR, RL, and RR).  We
denote, {\it e.g.}, $\sigma_{LL}$ as the cross section if both of the
incoming electrons are left-handed.

\begin{table}
\centering
\caption[]{
\label{table3}
\small
A comparison of the cross sections of SM processes for
different polarizations (LL, LR, RL, and RR) of the incoming electrons.
}
\medskip
\begin{tabular}{|l|l|}
\hline
Process $e^-e^- \to$  &   Relation among the $\sigma$'s \\
\hline
\hline
$e^-\nu_e W^-$   & $\sigma_{LL}>\sigma_{LR}=\sigma_{RL} >\sigma_{RR}=0$  \\
$e^-e^- Z$       & $\sigma_{LL}>\sigma_{LR}=\sigma_{RL} >\sigma_{RR}$  \\
$e^-e^- W^+ W^-$  & $\sigma_{LL}>\sigma_{LR}=\sigma_{RL} >\sigma_{RR}$  \\
$e^- \nu_e W^- Z$ & $\sigma_{LL}>\sigma_{LR}=\sigma_{RL} >\sigma_{RR}=0$  \\
$\nu_e \nu_e W^- W^-$ & $\sigma_{LL}>\sigma_{LR}=\sigma_{RL}=\sigma_{RR}=0$ \\
$e^- \nu_e W^- \gamma$ & $\sigma_{LL}>\sigma_{LR}=\sigma_{RL}>\sigma_{RR}=0$\\
$e^-e^- ZZ$   & $\sigma_{LL}>\sigma_{LR}=\sigma_{RL} >\sigma_{RR}$ \\
$e^-e^- H$  & $\sigma_{LL}>\sigma_{LR}=\sigma_{RL} >\sigma_{RR}$  \\
$e^- \nu_e W^- H$ & $\sigma_{LL}>\sigma_{LR}=\sigma_{RL} >\sigma_{RR}=0$  \\
$e^-e^- ZH$    & $\sigma_{LL}>\sigma_{LR}=\sigma_{RL} >\sigma_{RR}$ \\
\hline
\end{tabular}
\end{table}

\section{Summary}

We have calculated the Standard-Model cross sections for single and
double production of weak bosons, which are SM backgrounds in the
search for new physics in high energy $e^-e^-$ collisions. Single and
associated Higgs boson productions have also been calculated and can
give observable signals, although the cross sections are generally not
as large as those in $e^+e^-$ collisions.  We have also investigated
the possibility of observing strong $W^-_LW^-_L$ scattering, which
occurs through the weak isospin $I=2$ channel and is unique to
$e^-e^-$ collisions. We have developed certain kinematic cuts to
significantly reduce the $W^-_TW^-_T$, $W_T^-W_L^-$, $W^+W^-$ and $W^-Z$
backgrounds to the strong $W_L^- W^-_L$ signal in hadronic decay modes.
The $W^-Z$ background persists at
large $M_{WW}$, which makes the observation of strong $W^-_LW^-_L$
scattering difficult.  On the other hand, if doubly-charged Higgs
bosons exist, the $s$-channel enhancement in $W^-W^-$ final state
would be very substantial at an $e^-e^-$ collider.  This survey of
cross sections and processes should provide useful benchmarks for
serious studies of the potential of such a machine for new physics
discovery.

\section*{Acknowledgments}

We thank Pisin Chen for an informative discussion on the subject of
beamstrahlung. We also thank David Burke, Clemens Heusch, and Richard
Prepost for discussions.  This research was supported in part by the
U.S.~Department of Energy under Contract No.~DE-AC02-76ER00881 and in
part by the University of Wisconsin Research Committee with funds
granted by the Wisconsin Alumni Research Foundation. J.F.B. is
supported by a National Science Foundation Graduate Research
Fellowship. K.C. is supported by the U.S. D.O.E. grant
No.~DE-FG02-91ER40684.  T.H. is supported in part by the
U.S. D.O.E. grant No.~DE-FG03-91ER40674, and in part by a UC-Davis
Faculty Research Grant.

\newpage
\section*{Figure Captions}

\begin{enumerate}

\item\label{fig:a^3feyn}
Representative Feynman diagrams for single $W$ or $Z$ production in
$e^-e^-$ collisions.  In the case of $Z$ production, diagrams with the
interchange of final-state electrons must also be included.

\item\label{fig:a^3tot}
Total cross sections for single weak boson production in $e^-e^-$
collisions versus the CM energy $\sqrt s$.

\item\label{fig:a^4feyn}
Representative Feynman diagrams for $W^-W^-$ production in $e^-e^-$
collisions.  Diagrams with the interchange of identical final-state
particles ($\nu$'s and $W^-$'s) must also be included, omitting the
interchange of $W^-(k_1)$ and $W^-(k_2)$ in diagram (c).

\item\label{fig:a^4tot}
Total cross sections for vector boson pair production ($W^-W^+,\
W^-\gamma,\ W^-Z,\ W^-W^-,\ ZZ$) in $e^-e^-$ collisions versus the CM
energy $\sqrt s$.  The photon in the $W^-\gamma$ case has the
acceptance in Eq.~(3). The value $m_H=100$~GeV is assumed.

\item\label{fig:smfeyn}
Representative Feynman diagrams for the production of the
Standard-Model Higgs boson in $e^-e^-$ collisions.
The other diagrams can be obtained by the interchange $p_1 \leftrightarrow
p_2$.

\item\label{fig:smtot}
Cross sections for production of the Standard-Model Higgs boson in
$e^-e^-$ collisions (a)~versus $\sqrt s$ at $m_H=100$~GeV, (b)~versus
$m_H$ at $\sqrt s=0.5$~TeV.

\item\label{fig7}
The differential cross section $d\sigma/d m(b \bar b)$ versus the
invariant mass $m(b\bar b)$ of the $b\bar b$ pair in the process
$e^-e^- \to e^- e^- b\bar b$ at $\sqrt s=0.5$~TeV,
with the acceptance cuts of
Eqs.~(\ref{pTe cut}) and (\ref{pTb cut}).  The peak at $m(b\bar
b)\approx M_Z$ is due to $e^-e^-\to e^-e^- Z$ with $Z\to b\bar b$. The
signal due to a Higgs boson of mass $m_H=120$~GeV is illustrated.

\item\label{fig:pT} Distributions at $\sqrt s=2$~TeV in the transverse
momenta of vector bosons produced in the reactions $e^-e^-\to\nu\nu
W^-W^-, e^-e^-W^+W^-, e^-\nu W^-Z$: (a)~$p_T(VV) = \left|{\bf
p}_T(V_1)+{\bf p}_T(V_2)\right|$, (b)~$\Delta p_T(VV) = \left|{\bf
p}_T(V_1)-{\bf p}_T(V_2)\right|$.

\item\label{fig:invar} Invariant mass distributions of the weak boson
pairs produced in the reactions $e^-e^-\to\nu\nu W^-W^-, e^-e^-W^+W^-,
e^-\nu W^-Z$ after the acceptance cuts of Eqs.~(\ref{pT(VV)}) and
(\ref{Delta pT}) have been applied to enhance the strongly-interacting
$W^-W^-$ signal due to the exchange of a 1~TeV SM Higgs boson.

\item\label{fig:H--}
Representative Feynman diagrams for the production of $H^-$ and
$H^{--}$ in $e^-e^-$ collisions.  Diagrams with the interchange of
identical final-state particles ($\nu$'s and $H^-$'s) must also be
included.

\item\label{fig:MWWdoubly}
The distribution in the $W^-W^-$ invariant mass for $e^-e^-\to
\nu_e\nu_e W^-W^-$ including the contribution of a doubly-charged
Higgs boson of mass $M(H^{--})=0.2$ or 0.3~TeV.

\end{enumerate}

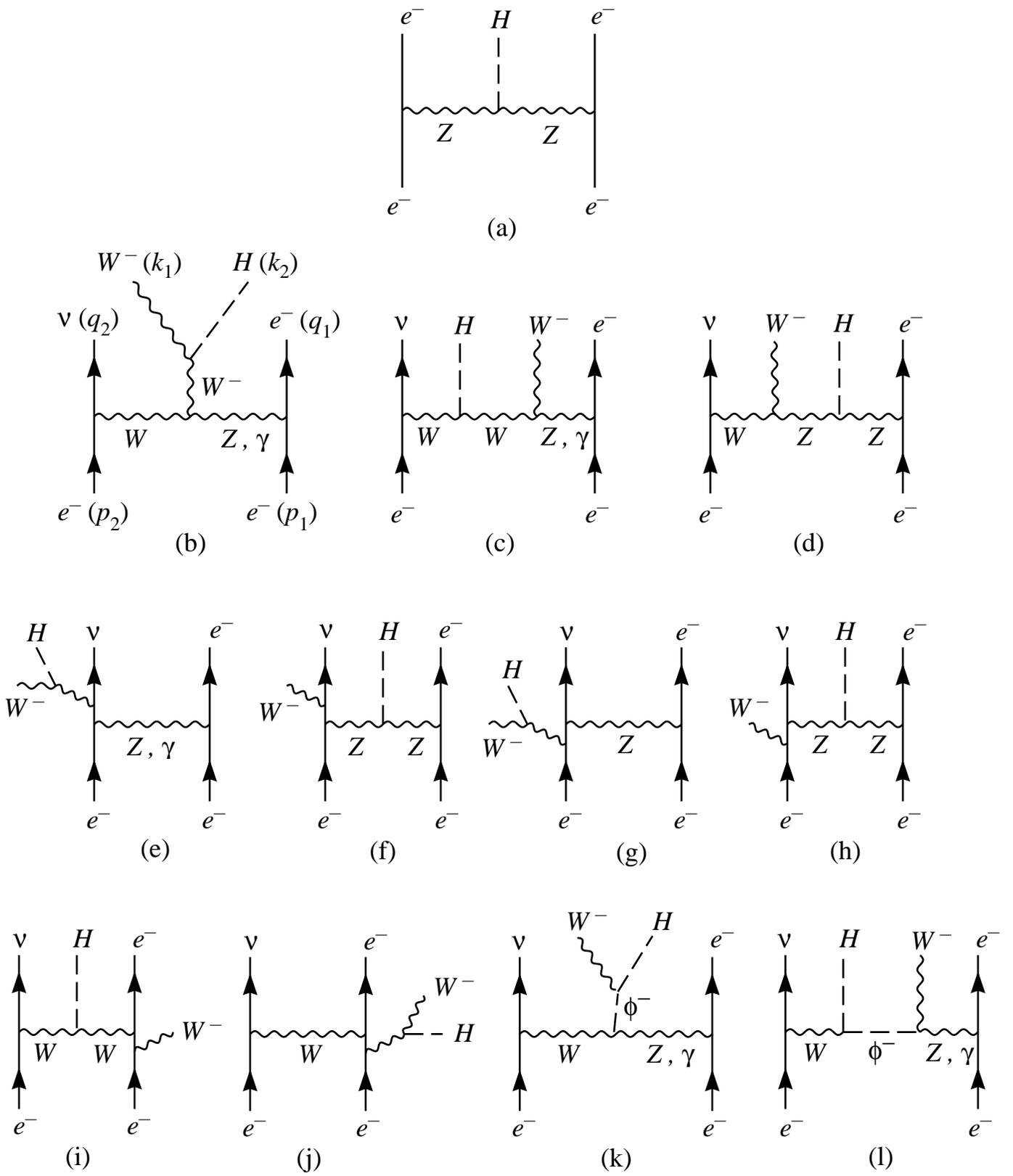
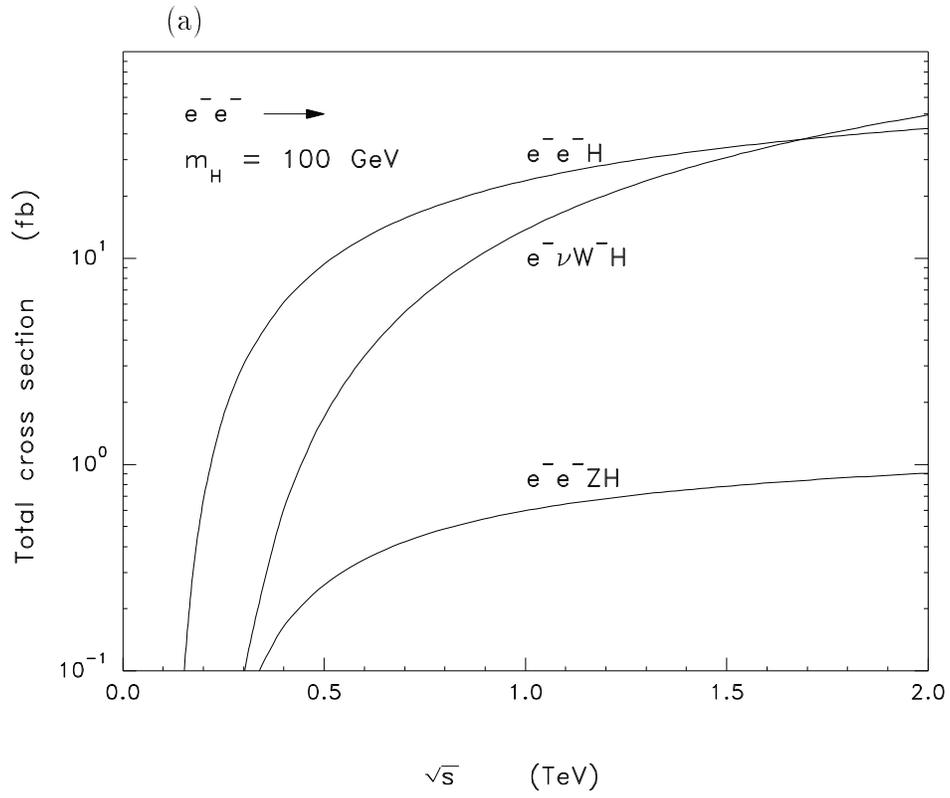
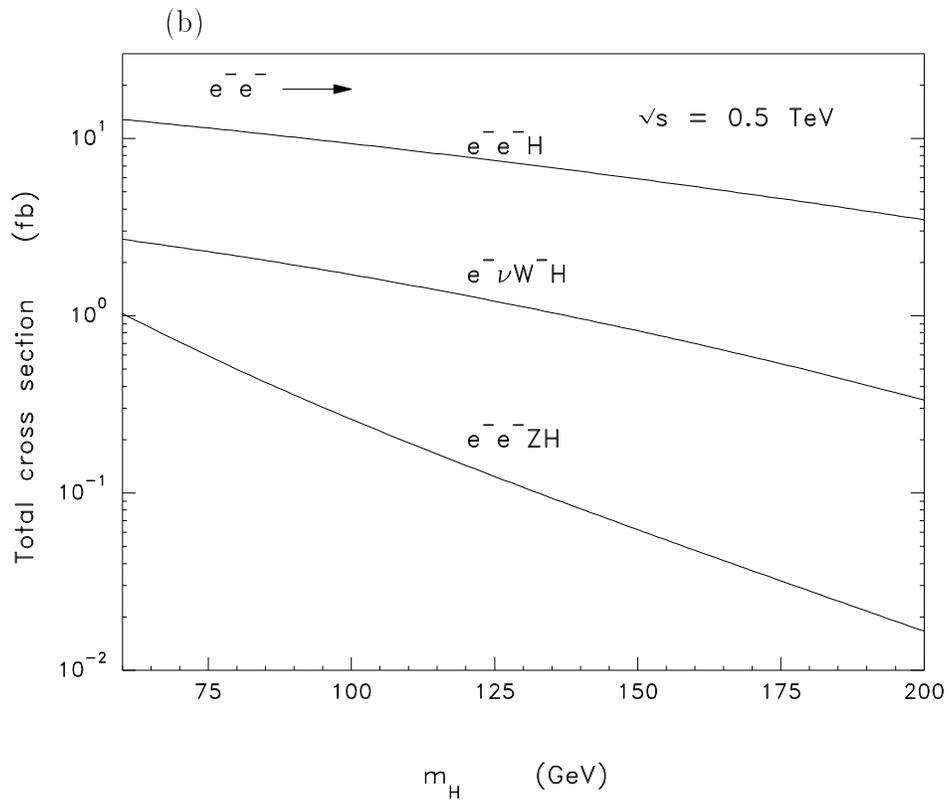
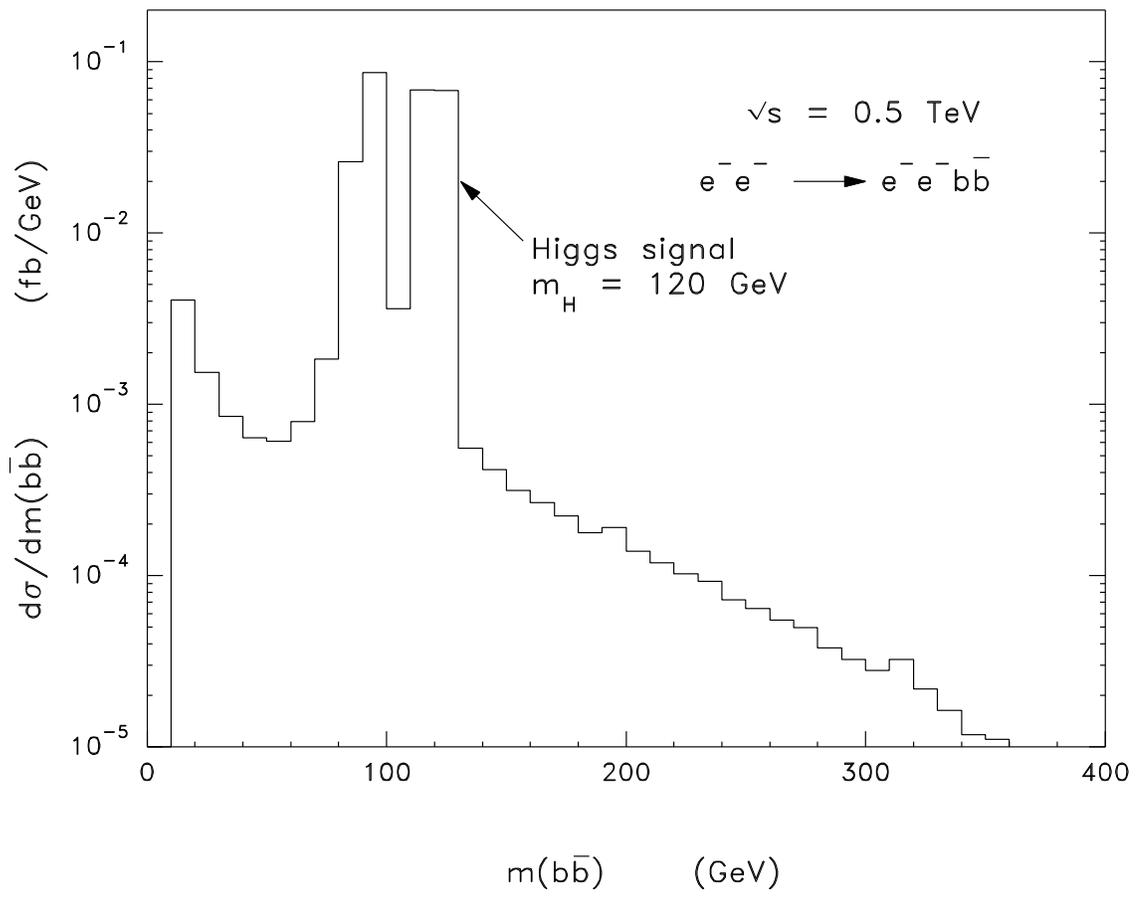
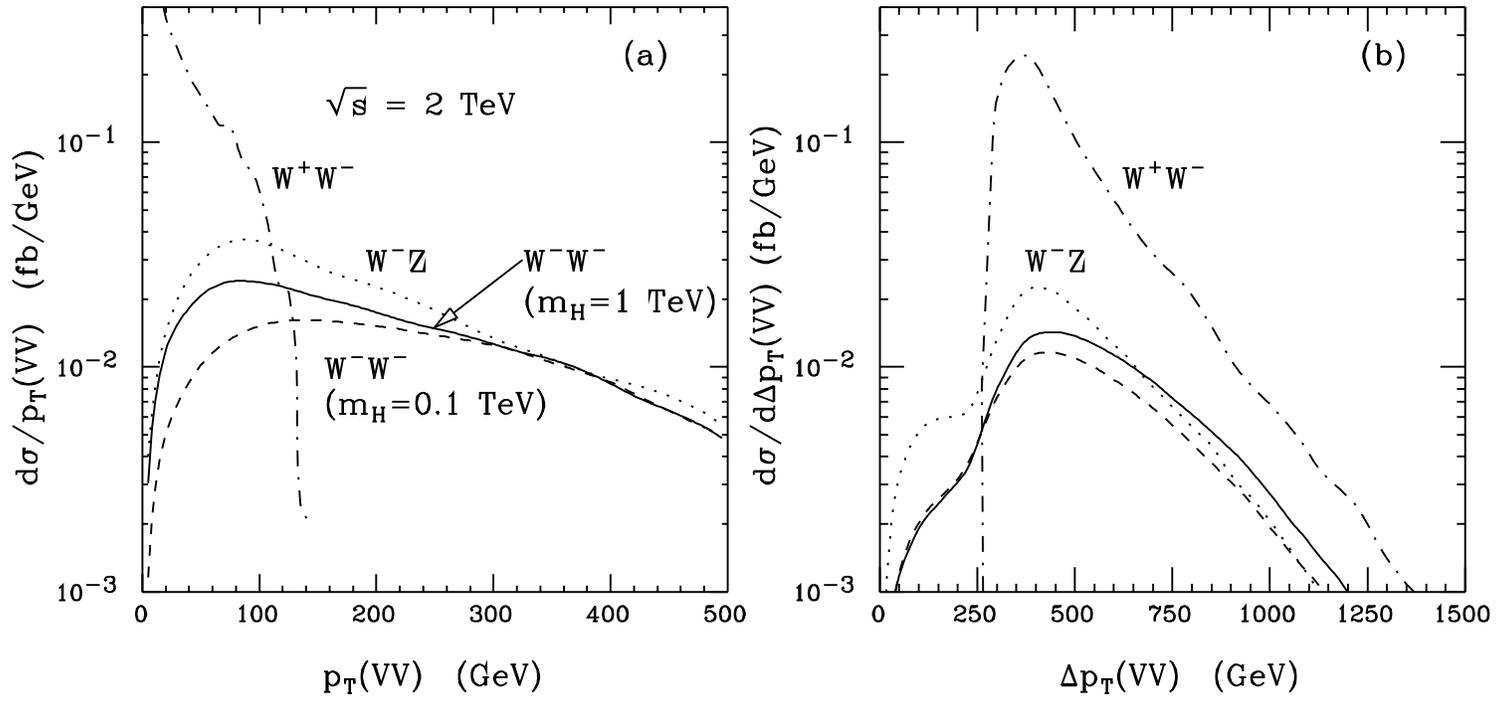
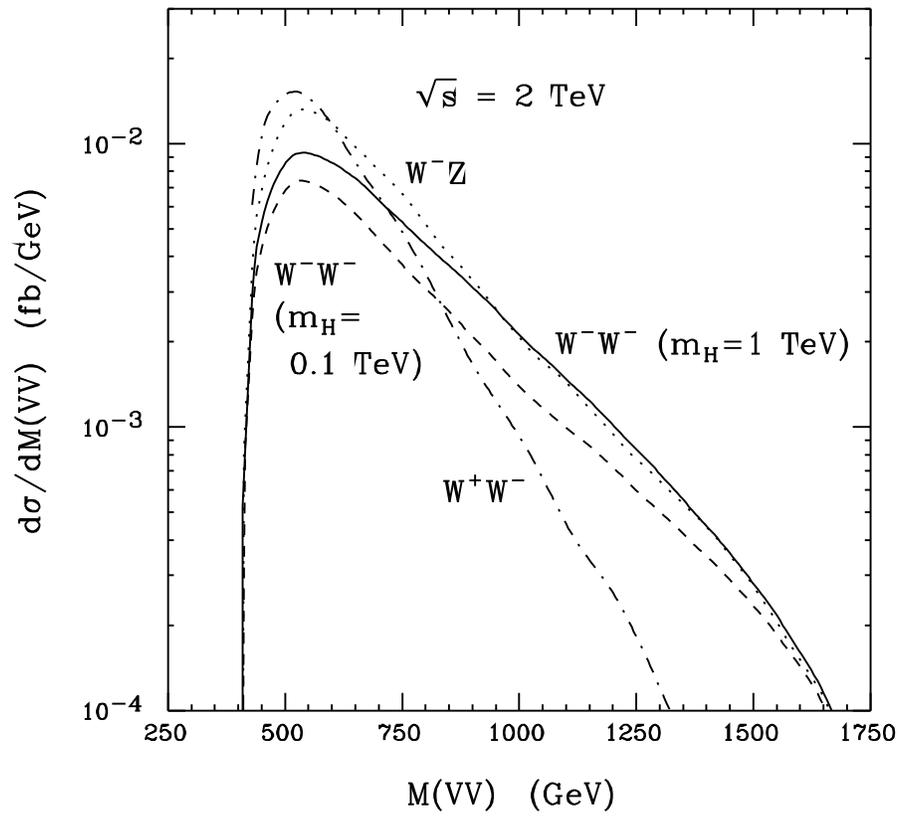
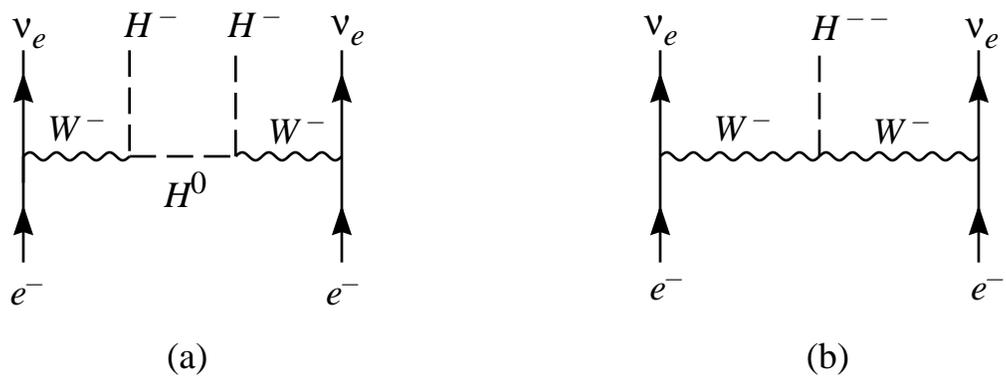
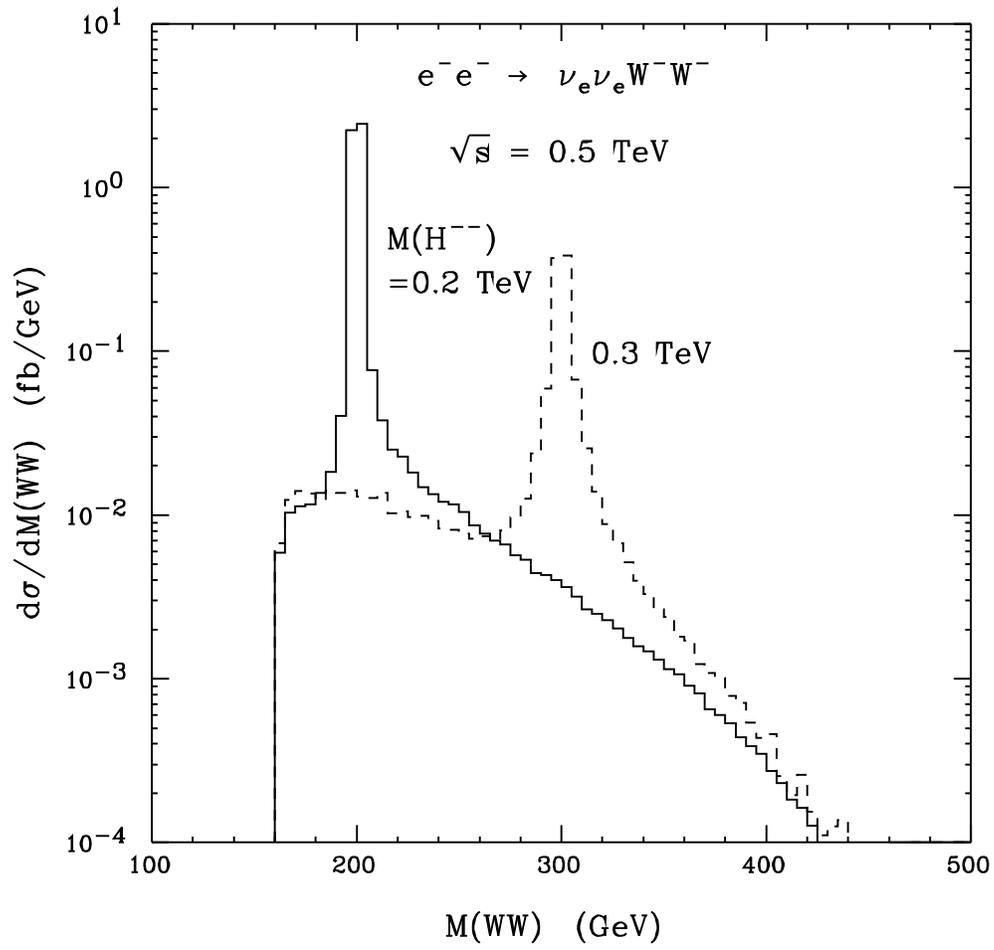
\end{document}